\documentclass[showpacs,amsmath,amssymb,onecolumn,superscriptaddress,notitlepage,preprintnumbers,pra]{revtex4-2}

\usepackage[dvips]{graphicx}
\usepackage{amsmath,amssymb,amsthm,mathrsfs,amsfonts,dsfont}
\usepackage{multirow}
\usepackage{array}
 \usepackage{qcircuit}

\usepackage{dcolumn}
\usepackage{bm}
\usepackage{amsmath}
\usepackage{amssymb}
\usepackage{braket}
\usepackage{physics}
\usepackage{amsthm}
\usepackage{natbib}
\usepackage{mathtools}
\usepackage{diagbox}
\usepackage[utf8]{inputenc}
\usepackage[english]{babel}
\usepackage{scalerel}[2014/03/10]
\usepackage{stackengine}
\usepackage{algorithm}
\usepackage{relsize}
\usepackage{xcolor}
\usepackage[noend]{algpseudocode} 
\usepackage[bottom]{footmisc}
\usepackage[flushleft]{threeparttable}
\usepackage{mdframed}
\usepackage{tikz}
\usepackage{listings}
\definecolor{maroon}{cmyk}{0,0.87,0.68,0.32}
\usetikzlibrary{backgrounds,fit,decorations.pathreplacing,calc,shapes.geometric}
\usetikzlibrary{positioning}
\usetikzlibrary{matrix}
\tikzset{set/.style={draw,circle,inner sep=0pt,align=center}}
\tikzstyle{int}=[draw, minimum size=3cm]

\usepackage{hyperref}
\hypersetup{colorlinks=true, linkcolor=blue, citecolor=blue, urlcolor=black }

\renewcommand{\tr}{\textup{Tr}}

\algrenewcommand\algorithmicrequire{\textbf{Input:}}
\algrenewcommand\algorithmicensure{\textbf{Output:}}

\makeatletter
\newcommand{\algmargin}{\the\ALG@thistlm}
\makeatother
\newlength{\whilewidth}
\algdef{SE}[parWHILE]{parWhile}{EndparWhile}[1]
  {\parbox[t]{\dimexpr\linewidth-\algmargin}{%
     \hangindent\whilewidth\strut\algorithmicwhile\ #1\ \algorithmicdo\strut}}{\algorithmicend\ \algorithmicwhile}%
\algnewcommand{\parState}[1]{\State%
  \parbox[t]{\dimexpr\linewidth-\algmargin}{\strut #1\strut}}

\theoremstyle{definition}

\newcounter{nicebox}
\newenvironment{nicebox}[1][]{%
 \stepcounter{nicebox}%
  \ifstrempty{#1}%
  {\mdfsetup{%
    frametitle={%
       \tikz[baseline=(current bounding box.east),outer sep=0pt]
        \node[anchor=east,rectangle,fill=blue!20]
        {\strut Box~\thenicebox};}}
  }%
  {\mdfsetup{%
     frametitle={%
       \tikz[baseline=(current bounding box.east),outer sep=0pt]
        \node[anchor=east,rectangle,fill=blue!20]
        {\strut Box~\thenicebox:~#1};}}%
   }%
   \mdfsetup{innertopmargin=10pt,linecolor=blue!20,%
             linewidth=2pt,topline=true,
             frametitleaboveskip=\dimexpr-\ht\strutbox\relax,}
   \begin{mdframed}[]\relax%
   }{\end{mdframed}}

\begin{document}

\title{A Herculean task: Classical simulation of quantum computers}

\author{Xiaosi Xu}
\affiliation{Graduate School of China Academy of Engineering Physics, Beijing 100193, China}

\author{Simon Benjamin}
\affiliation{Department of Materials, University of Oxford, Parks Road, Oxford OX1 3PH, United Kingdom}
\author{Jinzhao Sun}
\affiliation{QOLS, Blackett Laboratory, Imperial College London, London SW7 2AZ, United Kingdom}
\author{Xiao Yuan}
\affiliation{Center on Frontiers of Computing Studies, Peking University, Beijing 100871, China}
\affiliation{School of Computer Science, Peking University, Beijing 100871, China}

\author{Pan Zhang}
\affiliation{
 CAS Key Laboratory for Theoretical Physics, Institute of Theoretical Physics, Chinese Academy of Sciences, Beijing 100190, China
}
\affiliation{School of Fundamental Physics and Mathematical Sciences,
Hangzhou Institute for Advanced Study, UCAS, Hangzhou 310024, China}
\affiliation{International Center for Theoretical Physics Asia-Pacific, Beijing/Hangzhou, China}

\begin{abstract}

In the effort to develop useful quantum computers, simulating quantum machines with conventional computing resources is a key capability. Such simulations will always face limits, preventing the emulation of quantum computers of substantial scale; but by pushing the envelope as far as possible through optimal choices of algorithms and hardware, the value of the simulator tool is maximized. This work reviews the state-of-the-art numerical simulation methods, i.e., the classical algorithms that emulate quantum computer evolution under specific operations. We focus on the mainstream state-vector and tensor-network paradigms, while briefly mentioning alternative methods. Moreover, we review the diverse applications of simulation across different facets of quantum computer development, such as understanding the fundamental difference between quantum and classical computations, exploring algorithm design spaces for quantum advantage, predicting quantum processor performance at the design stage, and characterizing fabricated devices efficiently for fast iterations. 
This review complements recent surveys on today's tools and implementations; here, we seek to acquaint the reader with an essential understanding of the theoretical basis of classical simulations,  detailed discussions on the advantages and limitations of different methods, and the demands and challenges arising from practical use cases. 

\end{abstract}

\maketitle

\section{Introduction}

The invention of classical computers is undoubtedly one of the most significant breakthroughs in human history, and has brought revolutionary change to almost every aspect of life. Despite the remarkable success, classical computers are limited by memory and run time for many vital problems. On the other hand, quantum computers are believed to be able to bypass this scaling issue in various important cases and consequently there is the prospect that quantum processors will surpass classical systems in certain tasks. This concept is known as ``quantum advantage". Recent advancements in quantum hardware have resulted in the creation of quantum processors with hundreds of qubits, and there have been multiple claims that quantum advantage has been achieved with various types of quantum devices~\cite{arute2019quantum, zhong2020quantum,wu2021strong}. These achievements signify the beginning of the era of noisy intermediate-scale quantum (NISQ) devices.

In the efforts to push the field forward, 
classical computations play an essential role in assisting and accelerating the development of quantum computers. One can distinguish two basic categories:
\begin{enumerate}
    \item Classical simulation is a fundamental aspect of understanding and design of quantum hardware, frequently serving as the only means to validate these quantum systems;
    \item Quantum algorithms can be implemented on classical simulators for testing and verification purposes while full-fledged quantum computers are beyond reach. Classical simulators emulate the behavior of quantum computers and simulate the processes as if running on a real quantum device.
\end{enumerate}
Despite the development of numerous efficient classical simulators, there is a persistent need to simulate larger quantum systems. As quantum hardware has progressed to the point where it is no longer amenable to classical simulation\cite{arute2019quantum,wu2021strong,zhu2022quantum}, advancements in simulation capabilities that allow for the modeling of larger subsystems can offer valuable insights into many-body interactions, resulting in more accurate predictions of the performance of quantum algorithms and the behavior of quantum systems themselves.

Furthermore, it is important to continually push the limits of classical simulation, in order to understand the boundary between classical and quantum computations and to define the thresholds for practical quantum advantage.  

This review examines practical use cases in quantum computer development that drive the demand for faster classical simulation of larger quantum systems, ranging from quantum processor design~\cite{kyaw2021quantum,nickerson2013topological,bourassa2021blueprint}, hardware validations~\cite{villalonga2019flexible,mccaskey2018hybrid}, and characterizations~\cite{ganzhorn2020benchmarking,benedetti2019generative}, to quantum algorithm design and validation~\cite{huang2022provably,clinton2021hamiltonian,lu2021algorithms,wang2021variational}, and quantum advantage demonstrations~\cite{arute2019quantum, zhong2020quantum,wu2021strong}. 

As different use cases focus on different aspects of the quantum system, a large variety of classical methods exist that cannot be fully described in this single review. This review prioritizes the most general and widely-accepted circuit-based simulation methods for digital (versus analog) quantum computation, while others aimed for specific hardware including adiabatic simulation~\cite{albash2018adiabatic} and continuous variables~\cite{braunstein2005quantum} are beyond the scope of this review.  

There are several targets for classically simulating quantum computers. A seemingly natural choice is obtaining full information content of the final quantum state $|\psi\rangle$; however,  due to the exponential computational cost and memory requirement, this is affordable only for circuits with a small number of qubits. In many application scenarios, storing the entire state is unnecessary. For example, for stabilizer circuits that are commonly used in quantum error correction, one can convert the final-state-finding problem to the update of the set of Pauli operators that stabilize the state, using computational resources scaling polynomially to the number of qubits. In variational quantum eigensolver (VQE)~\cite{cerezo2021variational} and quantum approximate optimization algorithm (QAOA)~\cite{farhi2014quantum} simulations, as only the energy expectation is needed, circuits can be computed using, e.g., tensor network contractions with very limited storage~\cite{chen2018classical,huang2020classical}. In fact, many simulation methods have purpose-driven designs, and thus are appropriate for some specific application regime. 

For most application regimes, the different simulation methods can be categorized into four groups: exact and approximate, noiseless and noisy methods. As the names suggest, exact methods output accurate computation results, while approaches to get approximate results also exist with the benefit of cost reduction. Investigations into the in-principle performance of ideal hardware will employ a noiseless circuit, while noise should be considered in a more realistic scenario. 
Table 1 summarizes  the important measures for different classical simulation methods.

\begin{table}[ht]
\begin{threeparttable}\centering
\renewcommand{\arraystretch}{1.4}
\caption{Comparison between different simulation methods, where $N$ is the number of qubits, $m$ is the number of gates, $\chi$ is the bond dimension of MPS, $m_T$ is the number of $T$ gates in the circuit, and $W$ denotes the tree-width of the graph associated with the quantum circuit.}
\begin{tabular}{|m{0.8cm}<{\centering}|m{1.2cm}<{\centering}|m{1.8cm}<{\centering}|m{1.8cm}<{\centering}|m{1.6cm}<{\centering}|m{1.6cm}<{\centering}|m{3cm}<{\centering}|} 
\hline
\multicolumn{2}{|l|}{Methods}                & Memory & Run time & Approx. or exact & Noiseless or noisy & Application regime \\ 
\hline
\multirow{2}{0.7cm}{Full state} & State-vector   &   worst $O(2^N)$ & worst $O(m2^N)$ & Exact & Noiseless\tnote{\#} & General, good for small circuits\tnote{*} \\ 
\cline{2-7} & Density-matrix &  worst $O(2^{2N})$ & worst $O(m2^{2N})$ & Exact & Both &  General, good for small circuits\tnote{+} \\ 
\hline
\multicolumn{2}{|l|}{MPS state/MPO} &$O(N\chi^2)$   & $O(N\chi^6) $&Approx.  & Noisy& General, good for shallow circuits  \\ 
\hline
\multicolumn{2}{|l|}{Tensor network}  &On demand  &$O(e^{W})$ &Both  &Both & General, good for shallow circuits \\ 
\hline
\multicolumn{2}{|l|}{Stabilizer}  & $O(e^{m_T})$ & $O(e^{m_T})$ & Approx. & Both & Circuits dominated with Clifford gates, particularly in QEC \\ 
\hline
\end{tabular}
\begin{tablenotes}
\item[\#] State-vector simulators can also be used to simulate noisy circuits to get an approximate result with the Monte Carlo method.
\item[*] Circuits with $N>32$ with the state-vector simulator should generally run on an HPC server.
\item[+] Circuits with $N>16$ with the density-matrix simulator should generally run on an HPC server.
\end{tablenotes}
\end{threeparttable}
\end{table}

\section{Simulation methods}

\subsection{Full state}
The straightforward approach to simulate a quantum circuit is to record the full information of the quantum state and the changes as it evolves. Methods of full-state simulation include exact representations such as state vector and density matrix, and approximate ones such as matrix product state (operator) and neural network.

\subsubsection{Exact methods}   

In the state-vector formalism, the state of an $N$-qubit quantum register is represented by a $2^N$-dimensional vector of complex values. Performing universal quantum computation on a digital quantum computer involves a sequence of unitary operations applied to a quantum state.  The full-state simulator, or Schr\"{o}dinger simulator, takes the brute-force method to simulate a quantum circuit, where the entire vector of a quantum state is recorded and updated during the simulation process. The update of a quantum state can be done efficiently for gates in the Clifford group, as suggested by the Gottesman-Knill theorem. For example, for X and CNOT gates, one only needs to swap the amplitudes. However, the realization of universal quantum computation requires an additional non-Clifford gate, the common choice of which is the T gate. In the worst case with arbitrary unitary gates, the time and memory cost of a full-state simulation scale exponentially with the number of qubits. Typically $16$ bytes are needed per complex number and thus an $N$-qubit full-state vector in a naive implementation requires $2^{N+4}$ bytes, equivalent to $2^{N+4-30}$ GB of memory needed to perform such simulation. As a result, a personal laptop equipped with $16$ GB of memory will be capable of simulating up to about $30$ qubits.

\begin{nicebox}[Quantum gates, channels and circuits]

\textbf{The matrix form of some important quantum gates:}

$\mathbf{X}=\begin{bmatrix}
        0 & 1\\
        1 & 0\\
    \end{bmatrix}$,\,\,\,
$\mathbf{H}=\frac{1}{\sqrt{2}}\begin{bmatrix}
        1 & 1\\
        1 & -1\\
    \end{bmatrix}$,\,\,\,
$\mathbf{S}=\begin{bmatrix}
        1 & 0\\
        0 & i\\
    \end{bmatrix}$,\,\,\,
$\mathbf{T}=\begin{bmatrix}
        1 & 0\\
        0 & e^{i\pi/4}\\
  \end{bmatrix}$,\,\,\, 
$\mathbf{CNOT}=\begin{bmatrix}
        1 & 0 & 0 & 0\\
        0 & 1 & 0 & 0\\
        0 & 0 & 0 & 1\\
        0 & 0 & 1 & 0\\
    \end{bmatrix}$,\,\,\,
$\mathbf{CZ}=\begin{bmatrix}
        1 & 0 & 0 & 0\\
        0 & 1 & 0 & 0\\
        0 & 0 & 1 & 0\\
        0 & 0 & 0 & -1\\
    \end{bmatrix}$.

\vskip 10pt

\textbf{Clifford gates:} quantum gates that belong to the Clifford group, which is a group of unitary operators that can map Pauli gates to Pauli gates.

\vskip 10pt

\textbf{Kraus channel:} 
a quantum channel that is described by Kraus representation, where the state $\rho$ is evolved under $\xi(\rho)=\sum_{k}K_k\rho K_k^\dag$. The $\{K_k\}$ are called Kraus operators; and $\xi$ is trace preserving if $\sum_{k}K_k^\dag K_k=1$.

\vskip 10pt

\textbf{Quantum circuit:} Best illustrated by an example; the circuit below generates a GHZ state.
\vskip 5pt

    \centering\begin{tikzpicture}[thick]

    \tikzstyle{operator} = [draw,fill=white,minimum size=1.5em] 
    \tikzstyle{phase} = [fill,shape=circle,minimum size=5pt,inner sep=0pt]
    \node at (0,0) (q1) {$|0\rangle$};
    \node at (0,-1) (q2) {$|0\rangle$};
    \node at (0,-2) (q3) {$|0\rangle$};
    %
    \node[operator] (op11) at (1,0) {H} edge [-] (q1);
    \node[operator] (op21) at (1,-1) {H} edge [-] (q2);
    \node[operator] (op31) at (1,-2) {H} edge [-] (q3);
    %
    \node[phase] (phase11) at (2,0) {} edge [-] (op11);
    \node[phase] (phase12) at (2,-1) {} edge [-] (op21);
    \draw[-] (phase11) -- (phase12);
    %
    \node[phase] (phase21) at (3,0) {} edge [-] (phase11);
    \node[phase] (phase23) at (3,-2) {} edge [-] (op31);
    \draw[-] (phase21) -- (phase23);
    %
    \node[operator] (op24) at (4,-1) {H} edge [-] (phase12);
    \node[operator] (op34) at (4,-2) {H} edge [-] (phase23);
    %
    \node (end1) at (5,0) {} edge [-] (phase21);
    \node (end2) at (5,-1) {} edge [-] (op24);
    \node (end3) at (5,-2) {} edge [-] (op34);
    %
    \draw[decorate,decoration={brace},thick] (5,0.2) to
	node[midway,right] (bracket) {$\frac{|000\rangle+|111\rangle}{\sqrt{2}}$}
	(5,-2.2);
    \end{tikzpicture}

\end{nicebox}

In a more realistic scenario where noise is considered, one can perform an exact simulation with a mixed-state simulator. In this case, the state is expressed in the form of a density matrix and evolves under Kraus channels. The complexity of mixed-state simulations scales O($m2^{2N}$) where $m$ is the number of quantum gates. 

With the demanding memory cost as the system size increases, full-state simulators are not capable of simulating large quantum circuits. On the other hand, some approximations can be made to enable simulations of larger quantum systems at the cost of accuracy or run time. One ubiquitous example is modeling noisy circuits using the Monte Carlo method with a pure-state simulator, where the aggregated result of an adequate number of trials can well-approximate the exact result. This method can be further optimized to reduce the computational redundancy by pre-analyzing the noisy circuit and recording the most commonly-appearing intermediate state under noise~\cite{li2020eliminating}. Besides, various efforts have been made at the algorithmic level to optimize the full-state simulation. These include an adaptive encoding scheme~\cite{de2019massively} based on the polar representation that can reduce the memory to store a quantum state, at the cost of increasing the run time to perform the encoding and decoding steps; and a low-rank decomposition method~\cite{chen_low-rank_2021} to simulate noisy quantum circuits with density-matrix evolution where the density matrix is decomposed into a low-rank matrix with the most important columns.

Other algorithmic optimizations include methods designed for special-purpose simulations~\cite{jones2020natgrad,kassal2008polynomial} or for circuits with a specific structure~\cite{bravyi2016improved,vidal2003efficient}.  Moreover, optimizations at the implementation levels are also employed in practice, including parallelization using multithread~\cite{de2019massively}, acceleration with GPUs~\cite{amariutei2011parallel}, simulation reordering~\cite{fatima2021faster}, etc.
                
\subsubsection{Approximate methods}\label{sec:approximate}

In many quantum applications of interest, evolved quantum states are highly-structured and of low Schmidt rank, leading to a significant reduction in space and time complexity when storing and evolving them~\cite{vidal2003efficient}. A widely used approach in quantum information and quantum many-body physics is \textit{matrix product states} (MPS)~\cite{orus2014practical,schollwock2011density} (also known as \textit{tensor train} in mathematics~\cite{oseledets2011tensor,cichocki2016tensor}), and its mixed-state analog is matrix product operator (MPO)~\cite{pirvu2010matrix}, which is closely related to the density matrix renormalization group (DMRG) method~\cite{white1992density}. The MPS method represents the state using the contraction of three-way tensors with size $2\chi^2$ (and matrices on the boundary), as depicted in Box 2.
\begin{nicebox}[Tensor Network Diagram]
\centering\includegraphics[width=0.9\hsize]{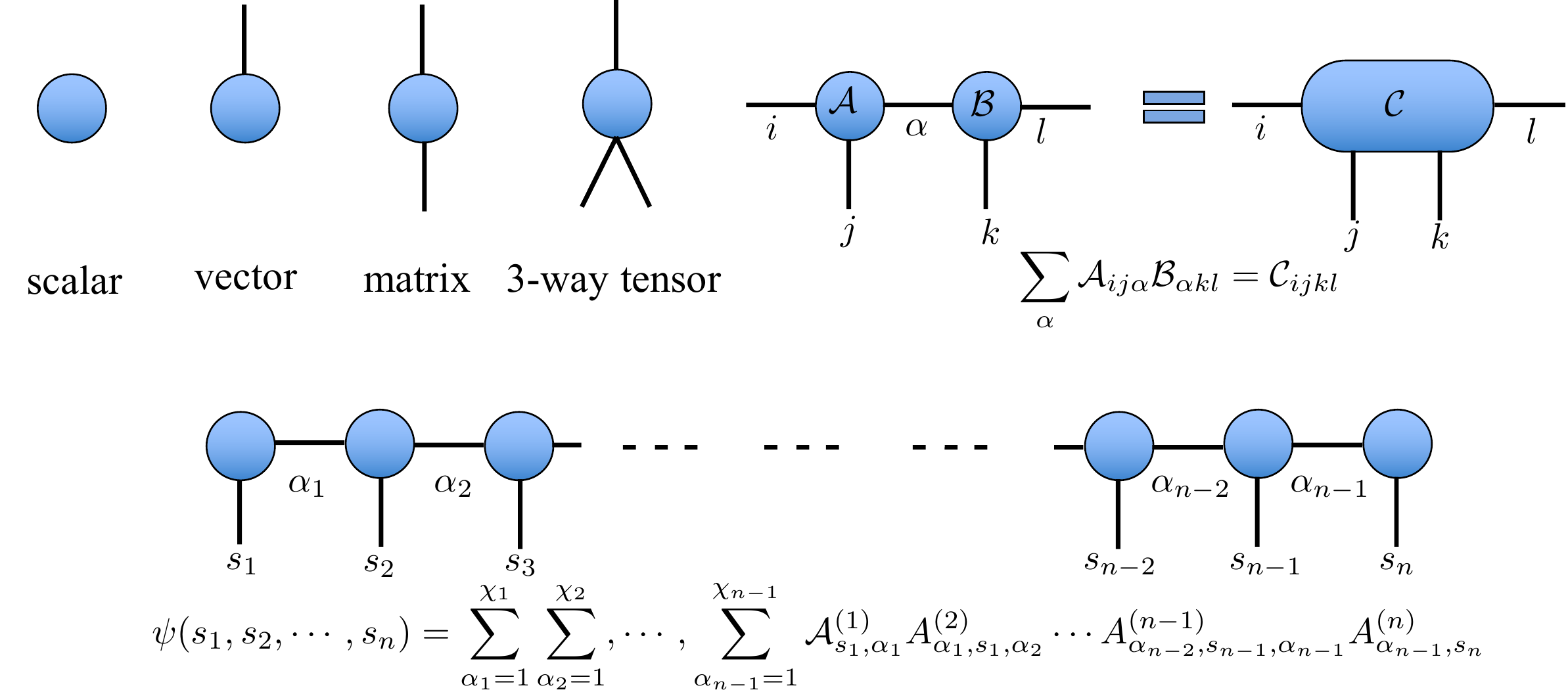}
\begin{flushleft}
Top: Diagram notations for tensors with different orders, and an example of contraction of two tensors.\\
Bottom: A matrix product state. 
$\{\mathcal A^{(i)}\}$ are three-way tensors (and matrices on the two boundaries), each of which has  dimension $d_i\chi^2$, with $d_i$ denoting the local dimension of physical indices $s_i$ and $\chi$ is the dimension of virtual indices $\alpha_1,\alpha_2,\cdots$.
\end{flushleft}
\end{nicebox}

In Box 2, \{$\mathcal A^{(i)}\}$ are three-way tensors (and matrices on the two boundaries), each of which has dimension $d_i\chi^2$, with $d_i$ denoting the local dimension of physical indices $s_i$ and $\chi$ is the dimension of virtual indices $\alpha_1,\alpha_2,\cdots$. As such, the MPS method is one of the simplest examples of tensor networks, with the difference that the state is evolved from the initial state to the final state, while the tensor network contraction method, which will be introduced in the next subsection, can employ general contraction orders. 

Compared with the state-vector method, the MPS representation reduces the number of parameters from $2^n$ to roughly $2n\chi^2$ (with $d_i=2$) at the cost of expressing only an approximation of the state with rather low entanglements~\cite{vidal2003efficient}.
It has been shown that the MPS method and its variant, the group MPS method, can efficiently simulate random circuits with controlled-Z gates and random Haar gates~\cite{zhou2020limits}. Further developments of the MPS simulation method~\cite{ayral2022density} manually close some qubits in the final state and combines the Markov chain Monte-Carlo sampling technique to investigate more challenging random circuits with the fSim gates, obtaining a cross-entropy benchmarking (XEB: a protocol introduced as a proxy to fidelity) that is similar to quantum hardware experiments on the Sycamore circuits~\cite{arute2019quantum} with simplified connection patterns.

Another way to approximate the state vector is by using neural networks~\cite{goodfellow2016deep,bishop2006pattern}. In contrast to the MPS method which explores intrinsic low-rank structures, neural network states exhibit a much higher rank of a tensor. However, it is difficult to apply unitary operators $U$ to the state neural network states $|\psi\rangle$ exactly. A solution is to choose neural networks that are easy to sample so that one can construct a loss function using samples for characterizing the distance between the old state $|\psi\rangle$ and new states $U|\psi\rangle$ after applying unitary operators, provided with sparse $U$ and $\ket{\psi}$, and then optimize the loss function to approximate the new state using neural networks. In Refs.~\cite{medvidovic2021classical,jonsson2018neural}, specific neural networks, restricted Boltzmann machines (RBM)~\cite{ackley_learning_1985,hinton_reducing_2006} have been employed to approximately represent and evolve the quantum state. It has been shown that it is convenient for simulating quantum circuits with two-qubit gates that have a diagonal matrix representation, e.g., controlled-Z gates and the gates in QAOA. However, minimizing the loss function introduces difficulties of non-convex optimizations, which heavily reduces the efficiency of the approach.

\subsection{Tensor networks}
\begin{nicebox}[Pictorial illustration of the tensor network of a quantum circuit]
\centering\includegraphics[width=0.8\hsize]{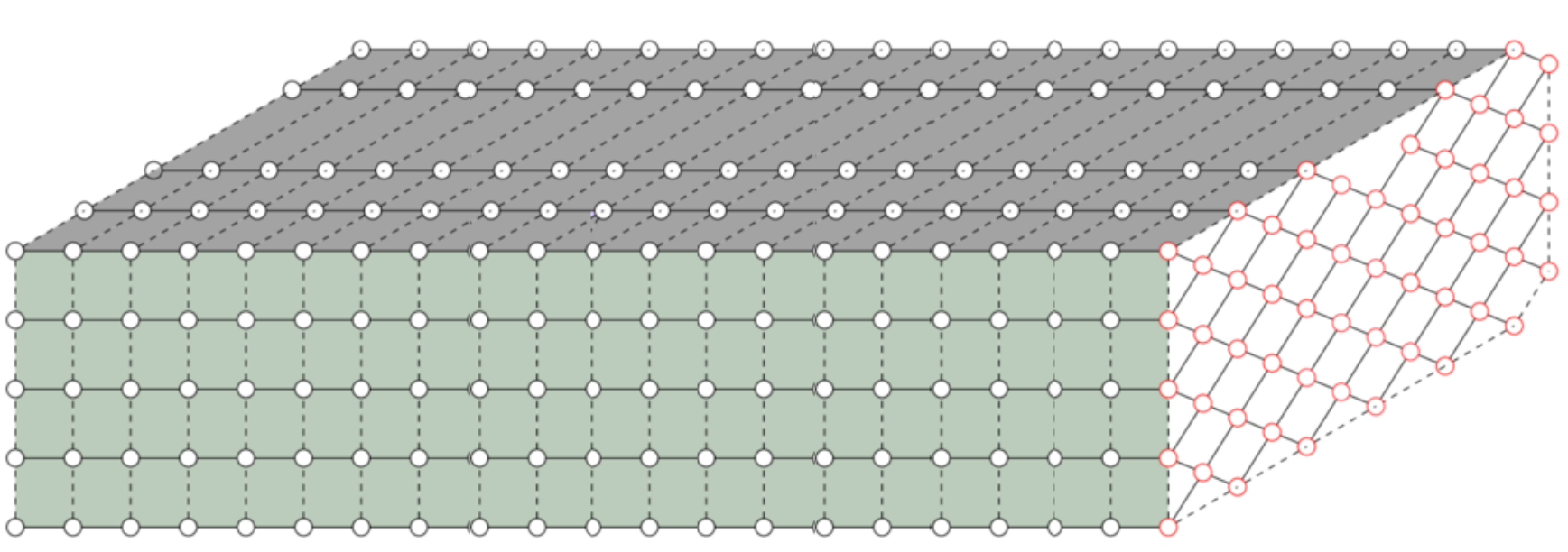}

\begin{flushleft}
If qubits are located on a 2-D layout, the evolution of the quantum circuit corresponds to a 3-dimensional tensor network. On the left-hand boundary of the tensor network, the tensors are initial states, which are usually product states, acting as a set of vectors on the left-hand side of the tensor network. On the right-hand side, the boundary corresponds to the final state of the quantum circuits. From the left boundary to the right boundary, there are quantum gates applying unitary transformations: the single-qubit gates are $2\times 2$ matrices and the two-qubit gates are order-4 tensors with dimension $2\times 2\times 2\times 2$. In other words, a quantum circuit is a subset of tensor networks, with the constraint that most of the tensors in the network are unitary. Thus, some of the circuit-simulation tasks can be converted to the tensor network contraction problem: If the purpose of the simulation is computing the amplitude of a given bitstring, the problem can be converted to a tensor network contraction problem with the other boundary (we will term it as ``right boundary'') of the tensor network being ``closed'' by a tensor product of vectors specified by the bitstring; If the target of simulation is a full state-vector, the right boundary of the tensor network is ``open'', or equivalently, having an identity matrix of size $2^n\times 2^n$ as the right boundary.
\end{flushleft}
\end{nicebox}

Tensor network contraction is a powerful method for simulating quantum circuits by representing the state or operator as a network of smaller tensors, as shown in Box 3. This reduces the memory and computational requirements, while also taking into account the specific structure of the circuit and qubit interactions. Additionally, the flexibility of the tensor network representation allows for the use of various contraction strategies, such as approximations and truncations, which can further increase simulation efficiency, as discussed in Sec.~\ref{sec:approximate}.

The use of tensor network contraction for simulating quantum circuits has its roots in the pioneering work of Markov and Shi in 2005 \cite{markov2008simulating}, which can be further reduced to node elimination under graphical representation. While efficient contraction methods exist for highly-structured tensor networks such as MPS and the multi-scale entanglement renormalization ansatz (MERA)~\cite{evenbly2011tensor}, the exact contraction of general tensor networks is known to be \#P-complete \cite{garcia2012exact}. Different simulation tasks specify different choices of boundary conditions for the corresponding tensor network, and can greatly impact the computational cost of contraction. An open boundary, also known as full-amplitude simulation, requires a space and time complexity exponential in the number of qubits. In contrast, a closed boundary, also known as single-amplitude simulation, may require less cost~\cite{gray2020hyper}. An intermediate option, known as the ``big-batch" boundary condition~\cite{pan2022simulation}, can compute a large number of correlated bitstring amplitudes by separating the final qubits into a closed part and an open part and placing a product of vectors on the closed qubits and a product of identity matrices on the open qubits. In some circuits, big-batch computation has a similar contraction cost to single-amplitude simulation and can be used to obtain full amplitudes by enumerating all the batches. When computing amplitudes for $m$ uncorrelated bitstrings, one can compute amplitudes for each bitstring individually by repeating contractions $m$ times. To improve efficiency, approaches such as the sparse state method~\cite{pan2022solving} and the multi-cache method~\cite{kalachev_recursive_2021} have been introduced, which can compute all the requested amplitudes simultaneously using just one contraction, avoiding the repetition of contraction for some parts of the tensor network shared by the computation for each bitstring.

The order in which tensors in a network are contracted can greatly affect the efficiency of the evaluation process. Different contraction sequences can result in varying levels of computational and storage demands. Finding the optimal contraction order for a given network can be a complex task, but various heuristic approaches have been proposed to tackle this challenge \cite{markov2008simulating,schutski2020simple}. For deep quantum circuits, the trivial contraction order, which starts from the initial state and progresses to the end state, is typically very close to the optimal order. However, for NISQ circuits, which are not that deep, the choice of the contraction order can have a significant impact on the computational cost.
To further enhance the capabilities of classical simulation, researchers have been investigating various high-performance computing techniques. However, the complex data dependencies between tensors can make it challenging to effectively parallelize the computation. Dynamical slicing methods, as introduced in \cite{chen2018classical,huang2021efficient}, solve this issue by splitting the target tensor network into multiple simpler networks with the same structure. This enables perfect parallelization and results in a significant boost in simulation efficiency. This approach has been widely adopted in various studies, such as Refs.~\cite{huang_alibaba_2019,huang2020classical,huang2021efficient,gray2020hyper,kalachev_recursive_2021}.
Tensor networks can be used not only to compute exact or approximate bitstring amplitudes and probabilities of the final state, but also to generate samples from a given distribution $P(s)$ with a specified fidelity or XEB~\cite{boixo2018characterizing}. Existing methods for this purpose include Frugal sampling~\cite{huang2020classical} and Markov-chain Monte-Carlo~\cite{pan2022solving, ayral2022density}. These methods leverage the capabilities of tensor networks to provide efficient sampling strategies for quantum circuits.

Furthermore, tensor networks provide a versatile framework for representing both closed and open systems~\cite{WBC15}, allowing for easy extension to density matrices and quantum channels~\cite{o2017density, huang_alibaba_2020}. However, their ability to simulate these systems exactly is still limited by the increasing resource demands.

\subsection{Stabilizers}

Although quantum circuits that can be simulated by the general-purpose classical simulators as described above are limited by size, there are some exceptions. The stabilizer circuit is one non-trivial example that has important applications in the field of quantum error correction. Stabilizer circuits in general are initialized to be stabilizer states, which are evolved under stabilizer operations and measured on a Pauli basis. As stabilizer operations are within the Clifford group, these circuits can be simulated efficiently by a classical computer. A couple of methods, by making use of the stabilizer tableau, have proven to be effective for the fast simulation of very large stabilizer circuits~\cite{PhysRevA.70.052328,Gidney2021stimfaststabilizer}. 

However, stabilizer circuits are not universal. To enable universal quantum computation needs an extension - the use of magic states is a canonical solution. As the cost of simulation scales exponentially with the number of magic states, when the circuits are dominated by Clifford gates, those circuits can still be simulated efficiently. There are mainly two classes of methods designed for near-stabilizer circuits, quasiprobability~\cite{PRXQuantum.2.010345,PhysRevLett.115.070501,PhysRevA.95.062337,PhysRevLett.123.170502} and stabilizer-rank~\cite{Bravyi2019simulationofquantum,PRXQuantum.2.010345,Qassim2019clifford,PhysRevA.99.052307}. The quasiprobability method for simulating quantum circuits is formalized by Ref.~\cite{PhysRevLett.115.070501}, with which an unbiased estimate of observables or the circuit outcomes can be obtained by sampling quasiprobability distributions over stabilizer states and operators. Quasiprobability allows the probability representation of a circuit element to be negative, while the negativity, the amount it deviates from a true probability distribution, serves as a major factor for the computation cost~\cite{PhysRevLett.115.070501}. Methods based on quasiprobability are natural to simulate noisy circuits; however, they are usually slower than stabilizer-rank-based methods. The stabilizer rank is defined as the smallest number of stabilizer states a quantum state can be decomposed into, which measures the simulation complexity. Determining the stabilizer rank is usually a hard problem~\cite{bravyi2016improved}, and an exact simulation can be costly. This motivates the work of Ref.~\cite{Bravyi2019simulationofquantum}, which proposes a sparsification method to obtain an approximation of the target state with a decomposition that contains fewer stabilizer states. This technique is further improved to have a reduced runtime in Ref.~\cite{Qassim2019clifford} and is generalized to qudits in Ref.~\cite{PhysRevA.99.052307} and the mixed-state case in Ref.~\cite{PRXQuantum.2.010345}.

\subsection{Other methods}
There also exist other classical simulation methods. One class of methods is based on the decision diagram~\cite{10.1007/978-3-540-87744-8_60,viamontes2009quantum,10.1007/978-3-319-08494-7_17,7163590,Zulehner2019AdvancedSO,hillmich2020just}, a graph-based approach that exploits redundancies of the quantum circuit and hence enables fast simulation of sizeable complex quantum circuits using limited resources. 
The state-of-the-art implementation of decision diagrams~\cite{Zulehner2019AdvancedSO} is comparable to that of MPS. 
Recently, a hybrid data structure, the tensor decision diagram~\cite{10.1145/3514355}, was proposed by combining the ideas of decision diagrams and tensor networks, integrating the flexibility of tensor networks while overcoming the memory bottleneck using decision diagrams in theory. As yet, only a proof-of-concept implementation has been done using the tensor decision diagram with up to 21 qubits~\cite{10.1145/3514355}. 

Another class of methods targets at non-Clifford circuits. A central ingredient is the computationally tractable (CT) states, for which the computational measurements on $\ket{\psi}$ can be simulated classically and the coefficient in this basis can be efficiently computed, i.e., (1)
it is possible to sample in $\mathrm{poly}(n)$ time from the distribution
$\mathrm{Pr}(x) = |\langle{x|\psi}\rangle|^2$ on the set of $n$-bit strings $x$ and (2) the coefficients $\langle{x|\psi}\rangle$ can be computed in $\mathrm{poly}(n)$ time.
Ref.~\cite{2009arXiv0911.1624V} shows that if the state $\ket{\psi}$ is CT and if  $U^{
\dagger}OU$ (circuit $U$ and measurement $O$)
is efficiently computable sparse, then the quantum computation of the sampling probability can be simulated classically.
Here, sparse quantum circuits include Pauli operators, local operators, and operators that can be written as a polynomial number of Toffoli gates.
                
There are also simulators specially designed for variational algorithms. The implementation of variational algorithms requires circuits to have the same structure but different gate parameters, and the circuit outputs can be reused across simulation runs. In Ref.~\cite{huang2021logical}, some techniques in artificial intelligence are used to convert a circuit into an arithmetic circuit that allows changing parameters and repeated measurements, enabling fast simulation compared with the general-purpose simulators.

\section{Applications}

Associated with the development of classical simulation methods there has been an emergence of a wide variety of software toolkits for public use, such as Qiskit~\cite{Qiskit}, ACQDP\cite{huang_alibaba_2019,huang_alibaba_2020,zhang2019alibaba}, Cotengra\cite{gray2020hyper}, Cirq~\cite{cirq}, QuEST~\cite{jones2019quest}, Yao.ji~\cite{YaoFramework2019}, Qjava~\cite{vizzotto2015qjava},
QETLAB~\cite{qetlab}. Those packages cover commonly-used programming platforms and have been used to explore and demonstrate diverse applications~\cite{huang_alibaba_2019,huang_alibaba_2020,zhang2019alibaba}. In this section, we give an overview of the key uses of classical simulations. These applications can generally be divided into two categories: the development of quantum hardware and the development of quantum algorithms. It is worth noting that some applications, such as quantum applications and hardware co-design, may fall into both categories.

\begin{figure}[!htb]
\centering\includegraphics[width=0.7\hsize]{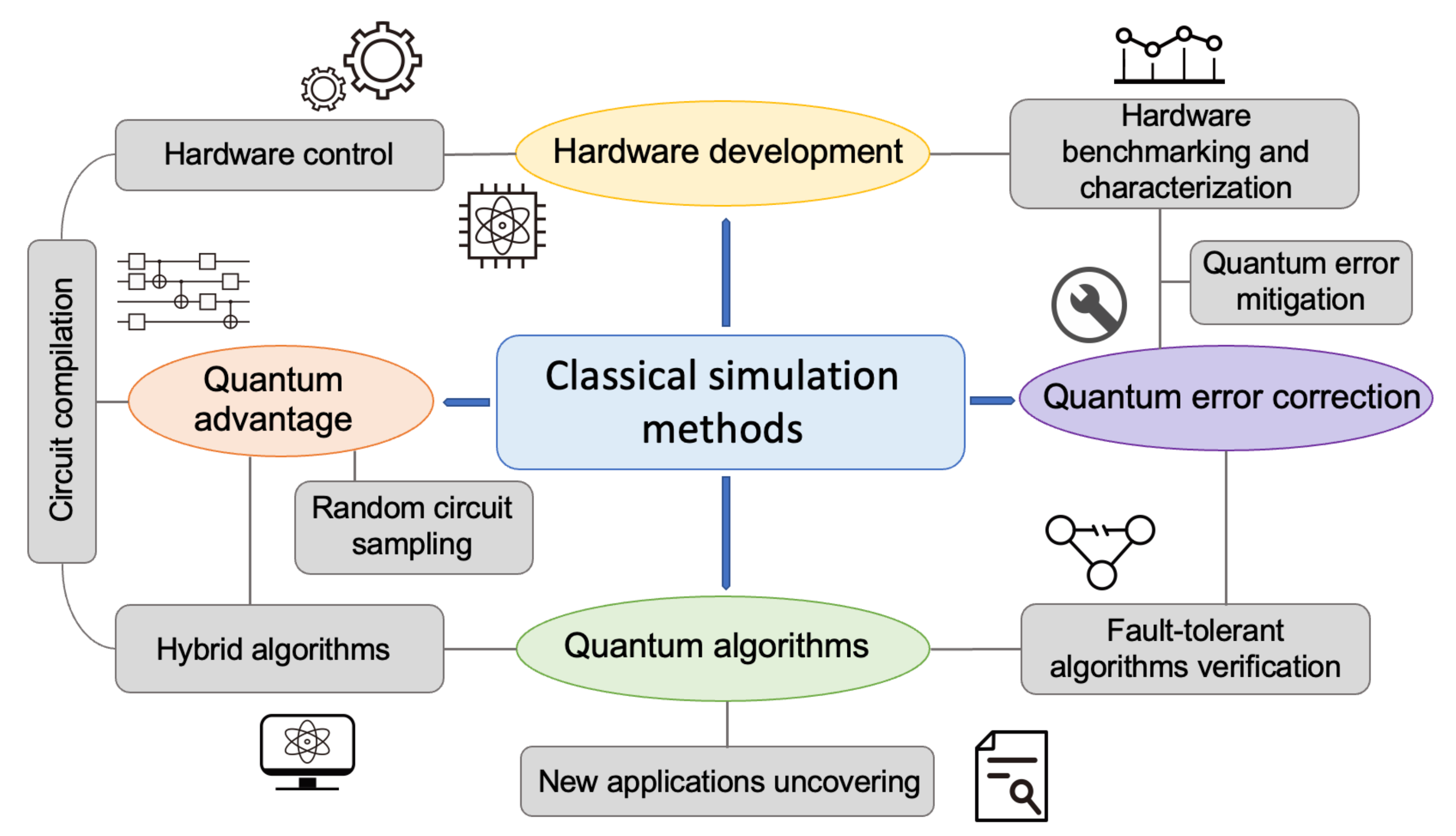}
\caption{Some key applications of classical simulations. Classical simulations can assist hardware development in modeling hardware control and benchmarks. In the field of quantum error correction and error mitigation, classical simulations are widely used to test the performance of error correction codes and mitigation strategies under different scenarios. Meanwhile, quantum algorithms are commonly simulated with classical methods for performance verification. Potential applications and problems with quantum advantage are also tested and demonstrated with classical simulations.}
\label{fig:app}
\end{figure}

\subsection{Quantum algorithm development}
Classical simulation is predominantly used in the development of quantum algorithms. Through the use of classical simulation methods, we can:
\begin{enumerate}
    \item understand the behavior of quantum algorithms under various conditions, identify potential sources of error and optimize the design and implementation of quantum algorithms;
    \item test and debug quantum algorithms before they are run on actual quantum hardware, saving time and resources in the development process.
    \item compare the performance of quantum algorithms to classical algorithms and evaluate the complexity of quantum algorithms in relation to the size of the input.
\end{enumerate}
These capabilities enable researchers to assess the feasibility of different quantum computing applications on various architectures and identify areas where quantum algorithms may be particularly advantageous. We will expand upon this in the following subsections.

\subsubsection{Quantum algorithm design and verification}

One challenge to the development of quantum technology is the lack of so-called killer applications. While certain well-known quantum algorithms, such as integer factoring and Hamiltonian simulation, have provable performance, those empirical quantum algorithms may require a platform to verify and validate if they actually work. Specific examples include an empirical algorithm to distinguish non-isomorphic graphs, with experiments to validate its performance in terms of run time and solution quality~\cite{huang_alibaba_2019}. Classical simulation methods are also commonly used to test their robustness under realistic circumstances, e.g., with noise or gate implementations.
Moreover, before algorithms are implemented on a real device to get a meaningful and reliable result, prototype demonstrations for particular applications are commonly conducted on classical computers to uncover research in new paradigms and raise interest across academia and industry for potential applications of quantum computers.

\subsubsection{Quantum advantage demonstration and validation}

A key question for the quantum computing field is when, and to what context, quantum computers will outperform conventional machines. Terms such as `quantum supremacy' (QS) and `quantum advantage' (QA) are typically used in this discussion. However these terms are often used inconsistently in different papers. QS is deemed to occur when {\it any} task can be performed on a quantum computer with profoundly lower resource costs (primarily, time) than a classical alternative. The task in question need not be {\it useful} and may indeed be defined solely for the QS demonstration. Somewhat confusingly, some authors use QA as a synonym for QS, finding the word `advantage' less problematic than `supremacy' in terms of its historical resonance. However, many other authors prefer to reserve the use of QA to refer to scenarios where the task at hand is actually useful, at least potentially, users beyond the quantum computing community.

Whether one considers QS or QA, the challenge drives the optimization of classical simulation methods, with the aim of verifying claim that the quantum system has fundamentally better performance. Validation is critical as new proposals may not have undergone thorough prior examination, and it is crucial to ensure that research is based on sound foundations and that results are reliable.

Random circuit sampling (RCS) is a leading proposal to demonstrate that a quantum computer can go beyond the capability of a conventional machine. A random quantum circuit is used to generate samples from its output distribution~\cite{aaronson2016complexity,boixo2018characterizing}. With a qubit count exceeding $50$, it is anticipated that simulating the full state of the system would require $16$ petabytes of memory, which is eight times more than that of the most powerful supercomputer at the time. The circuit structure is also specifically designed to achieve a large tree width, making it challenging to simulate using tensor network methods. This is thus a task well-suited to distinguishing between quantum and classical systems, but is not generally considered useful beyond that context -- the random nature of the circuits makes this so practically by definition. 

The initial demonstration of quantum supremacy was conducted by Google in 2019 \cite{arute2019quantum}. Since then, ongoing efforts have been made to develop classical simulation methods that challenge the validity of the claim. RCS has also become a benchmark for quantum hardware, as it not only takes into account qubit count, but also qubit quality, which can be characterized using XEB. The continuous advancements in quantum hardware \cite{wu2021strong, zhu2022quantum} also push the boundaries of quantum computing and pose new challenges for classical simulation methods to keep pace.

As an illustration of how classical methods contribute to this task, Box 4 provides further information on the various strategies used to implement the sampling problem using classical simulation methods.

\begin{nicebox}[Random circuit sampling and its verifications]
{\bf Sampling problem of quantum circuits}\\
In the experimental realization, the quantum hardware generates bitstring samples $\{\mathbf s\}$ by performing measurements on the states produced by applying the circuit $U$ composed of non-perfect quantum gates with randomly generated parameters. 
{Owing to noise, the underlying distribution $Q(s)$ of the samples is not the distribution $P_U(s)$ corresponding to the exact final state $|\psi\rangle$.}
This sampling problem can also be solved using classical algorithms. The most straightforward simulation method is the full-amplitude simulation by storing and evolving the state vector~\cite{boixo2018characterizing,pednault2019leveraging}. But it is not suitable for the sampling problem with a large number of qubits. The tensor network methods allow computing exact or approximate amplitudes for only part of the bitstrings that are further used for generating samples and are for the sampling problems with shallow circuits.\\

\noindent {\bf Verification of the sampling problem}\\
An important problem is how to characterize the quality of samples generated by quantum hardware or classical simulation. One seeks to estimate the distance between the generation distribution $Q(s)$ and the true distribution $P_U(s)$. In literature, the quality of samples is characterized using the cross entropy benchmark (XEB) for random circuits where the Porter-Thomas rule holds~\cite{boixo2018characterizing,arute2019quantum}, and is computed using the difference between cross-entropy of $Q(s)$ and $P_U(s)$ and the cross entropy between $Q(s)$ and the uniform distribution. It has been used for demonstrating meaningful gaps between quantum and classical systems ~\cite{arute2019quantum,wu2021strong,zhu2022quantum}. The exact value of XEB for large circuits requires one to make comparisons with classical predictions. 
\\

\noindent {\bf Tensor network methods for the sampling problem}\\
Initially, tensor network contraction was used for 
computing exact bitstring amplitudes (and probabilities) one by one~\cite{boixo2017simulation,chen2018classical} which were further used for generating samples via e.g. frugal sampling methods~\cite{markov2018quantum}.
Subsequently, it has been shown that a small group of $k$ correlated bitstring amplitudes can be computed with almost the same computational cost for computing a single bitstring amplitude~\cite{villalonga2019flexible}. The approach has been shown to be very efficient in solving sampling problems for 2D quantum circuits with controlled-Z gates~\cite{markov2018quantum,villalonga2019flexible,zhang2019alibaba,guo2019general}. 

A breakthrough in the tensor network approach for quantum circuit simulation is the balanced-partitioning-based approach for finding contraction orders~\cite{kourtis2018fast,gray2020hyper}. By combining the partitioning-based contraction-order finding and frugal sampling, researchers have proposed~\cite{huang2020classical} to generate some perfect samples by repeating the tensor network contraction many times and mixing with random bitstrings, so as to arrive at a set of samples with XEB close to Google's hardware, with a computation time of roughly $20$ days on a supercomputer.

If the objective is only obtaining a higher XEB (i.e., spoofing of XEB~\cite{gao2021limitations}), the computational cost can be heavily reduced. It was proposed~\cite{pan2021simulating} that a big-batch (e.g., of size $2^{21}$) of correlated bitstrings with exact bitstring probabilities can be computed efficiently and used for generating correlated samples with XEB much larger than Google's hardware XEB. This big-batch approach has also been implemented on a supercomputer~\cite{liu2021closing}, requiring only $304$ seconds. Notice that sampling from a big batch of correlated bitstrings gives correlated samples, which is significantly different from Google's experiments which produce uncorrelated samples. In~\cite{pan2022solving}, the spare-state method was proposed to obtain one million uncorrelated bitstring samples of the Sycamore circuits with estimated fidelity greater than Google's hardware XEB, using a single network contraction. 

\end{nicebox}

\subsubsection{Quantum error correction} 

As  discussed in the previous sections, stabilizer circuits can be  simulated efficiently and are widely used in estimating the performance of quantum error correction codes, such as  logical error rates and thresholds. The efficient simulatability is based on the assumption that noise sources can be described by Pauli operators drawn from some fixed distribution. The premise is good enough to provide conceptual guidance. However, a simple error model cannot capture the behavior of real quantum hardware, such as qubit idling, photon decay, flux noise, etc. Therefore, a series of works have been devoted to simulating the surface code under more realistic error models~\cite{tomita2014low,o2017density,huang2021logical,katsuda2022simulation} by using simulation techniques such as density-matrix simulation or a tensor network approach. However, the highest code distance is simulatable under the surface code architecture and a realistic error model is still limited. 

We should note that quantum error correction circuits and NISQ circuits have different parameter regimes in circuit size and depth. Generally, quantum error correction circuits will be much longer than usual NISQ circuits. This makes massive parallelization techniques developed for a tensor network approach much less beneficial, but provides more opportunities for developing new classical simulation techniques.

\subsubsection{Circuit compilation}

Efficient classical simulation of quantum circuits also enables the investigation of circuit compilation, specifically to compile a unitary or a quantum state with shallow-depth circuits consisting of local gates or others that satisfy certain hardware restrictions. This is essential for  accurate implementations of quantum algorithms  and hence is important for realizing quantum advantage using limited near-term quantum computers. 
For unitary compilation~\cite{Heya18,Khatri_2019,Sharma_2020}, the task is to optimize an operator $U(\vec\theta)$ with classical variables $\vec\theta$, so that it is close to the target unitary $V$. One can use $|\tr[U(\vec\theta)^\dag V]|^2$ (or others) as the cost function, which could be obtained via quantum simulators, and maximize it to compile $V$. Here the operator $U(\vec\theta)$ could be either a parametrized quantum circuit or the Hamiltonian evolution of the hardware, possibly with implementation restrictions for compilation with a real device. The compiled operator $U(\vec\theta)$ generally has a much shallower quantum circuit, which is more experimentally friendly and more noise-robust. 
One can also consider the compilation for specific quantum states, where, in this case, the cost function is replaced by $|\langle{\psi_0|U(\vec\theta)^\dag V|\psi_0}\rangle|^2$ for a certain initial state $\ket{\psi_0}$. The compilation algorithm has been tested for up to 20 qubits using a high-performance classical simulator~\cite{Jones2022robustquantum}.
Furthermore, classical simulators could be applied for compiling error correction circuits with shallower depth and hardware-friendly quantum gates. Refs.~\cite{Johnson17,Xiaosi21ErrorCorrection} have shown the compilation for the five- and seven-qubit codes. 
More efficient classical simulators are required for the circuit compilation of larger systems.

\subsection{Quantum hardware development} 
While quantum algorithm development may receive more attention, the development of quantum hardware is another vital and growing field that also benefits from classical simulation techniques. Classical simulations can be used to study a wide range of factors that affect the performance of quantum hardware, such as the effects of noise and decoherence, the optimization of quantum circuits, and the characterization of fabricated quantum devices. By using classical simulations, researchers can gain valuable insights into the behavior of quantum systems, test and debug quantum hardware before it is built for faster iteration, and identify potential challenges and opportunities for improving quantum hardware.

\subsubsection{Quantum hardware design}
In the development of quantum hardware, the chip design process necessitates a substantial amount of classical simulation work to evaluate whether a given superconducting circuit design (or other qubit platforms) can perform high-fidelity quantum operations. The exact diagonalization of the full circuit Hamiltonian becomes computationally intractable for large systems, highlighting the need for new methods to efficiently model these systems. Recent work has employed tensor network methods to model superconducting circuits spanning a Hilbert space as large as $15^{180}$~\cite{di2021efficient}. However, while it has been reported that it is feasible to study coherence time using this approach, it remains highly challenging to accurately model and predict quantum operation fidelities for systems involving a few dozen qubits.

\subsubsection{Hardware benchmarking and characterization} 

Another motivation for the ongoing drive to improve classical simulation methods is to aid in the validation of quantum hardware. One specific area of focus is the development of quality assurance techniques for quantum processors. Various metrics have been proposed and extensively used for benchmarking quantum processors, such as XEB~\cite{arute2019quantum} and quantum volume~\cite{PhysRevA.100.032328}. However, as the system size grows, the exponential computational resources required for these methods become increasingly demanding. In response, alternative perspectives have been developed to address this challenge, such as the use of XEB with Clifford circuits~\cite{chen2022linear}, which require only modest classical simulation resources. These efforts aim to provide more efficient and practical ways to evaluate the performance of quantum processors, and to aid in the ongoing advancement of quantum hardware.

\section{Challenges and outlook}

Even though quantum computers are anticipated to outperform classical computers ultimately, classical simulations are still indispensable at the current stage. We have witnessed the development and adoption of various classical simulation methods for quantum computers, and their wide applications in different tasks in developing quantum hardware and quantum algorithms. With the fast development of quantum computers, the need for more efficient classical simulation methods will also rise rapidly. 

The fundamental challenge in the classical simulation of quantum systems is the exponential scaling of the resources required as the quantum system size grows. This is because the quantum state is represented by a complex-valued wave function, which requires an exponentially large number of complex numbers to describe. As a result, it is generally very difficult to exactly simulate large quantum systems with classical computers.

There are a number of approximate methods that have been developed to overcome this difficulty and to allow for the classical simulation of large quantum systems. These methods include MPS/MPO, tensor network, and stabilizer techniques. However, several challenges arise in the approximate classical simulation methods themselves:
\begin{enumerate}
    \item Accuracy: One of the main challenges is finding methods that can accurately approximate the simulated quantum system's behavior, especially for systems that exhibit complex or highly correlated behavior.
    \item Scalability: Another challenge is to find scalable methods that can efficiently simulate larger quantum systems. Many approximate methods also suffer from exponential scaling, meaning they become impractical for larger systems.
    \item Efficient implementation: A third challenge is to develop efficient algorithms and approaches for implementing these approximate methods on classical computers. 
    \item Validation: A final challenge is to validate the accuracy of the approximate methods being used, as it may not always be possible to exactly simulate the quantum system and compare the results to the approximate solution.
\end{enumerate}

Yet another challenging task in the classical simulation of quantum systems is the development of efficient algorithms for simulating specific quantum systems or quantum algorithms. While there has been significant progress in this area, there is still much work to be done to fully understand the capabilities and limitations of classical simulation for specific types of quantum systems and algorithms. 

We hope this review will stimulate further development of classical simulation methods of quantum computers with new simulation methodologies, optimized implementations, and broader applications; such progress will be ever more valuable for benchmarking, verifying, and designing quantum computers and algorithms.

\begin{acknowledgments}
The authors thank Tyson Jones and Suguru Endo for early discussions on this review. 
\end{acknowledgments}

\bibliographystyle{unsrt}
\bibliography{bibliography.bib}

\end{document}